# Low Gain Avalanche Detectors with Good Time Resolution Developed by IHEP and IME for ATLAS HGTD project

Mei Zhao[a,b], Xuewei Jia[a,c], Kewei Wu[a,c], Tao Yang[a,c], Mengzhao Li[a,c], Yunyun Fan[a,c], Gangping Yan[d], Wei Wang [a,c], Mengran Li [a,c], Gaobo Xu[d], Mingzheng Ding [d], Huaxiang Yin [d], Jun Luo [d], Junfeng Li [d], Xin Shi[a,b], Zhijun Liang[a,b], João Guimarães da Costa[a]

[a] Institute of High Energy Physics, Chinese Academy of Sciences, 19B Yuquan Road, Shijingshan District, Beijing 100049, China
[b] State Key Laboratory of Particle Detection and Electronics, 19B Yuquan Road, Shijingshan District, Beijing 100049, China
[c] University of Chinese Academy of Sciences, 19A Yuquan Road, Shijingshan District, Beijing 100049, China
[d] Institute of Microelectronics, Chinese Academy of Sciences, Beitucheng West Road, Chaoyang District, Beijing 100029, China

**Abstract**

This paper shows the simulation and test results of 50 μm thick Low Gain Avalanche Detectors (LGAD) sensors designed by the Institute of High Energy Physics (IHEP) and fabricated by the Institute of Microelectronics of the Chinese Academy of Sciences (IME). Three wafers have been produced with four different gain layer implant doses each. Different production processes, including variation in the n++ layer implant energy and carbon co-implantation were used. Test results show that the IHEP-IME sensors with the higher dose of gain layer have lower breakdown voltages and higher gain layer voltages from capacitance-voltage properties, which are consistent with the TCAD simulation. Beta test results show that the time resolution of IHEP-IME sensors is better than 35ps when operated at high voltage and the collected charges of IHEP-IME sensors are larger than 15fC before irradiation, which fulfill the required specifications of sensors before irradiations for the ATLAS HGTD project.
**Keywords:** Low Gain Avalanche Detectors (LGAD), implantation dose, breakdown voltage, time resolution, charge collection

Email address: zhaomei@ihep.ac.cn (Mei Zhao)

January 21, 2022

# 1. Introduction

The High Luminosity Large Hadrons Collider (HL-LHC) will start in 2027 with a significant increase in instantaneous luminosity. As the instantaneous luminosity increases, the number of collisions in each bunch crossing (pile-up) increases. To separate collisions in limited space, the choice of solid-state timing detectors for ATLAS High Granularity Timing Detector (HGTD) project [2] are presently thin Low Gain Avalanche Detectors (LGAD) [3, 4], which have time resolution better than 50 ps. So far, the LGAD sensors have been developed by several silicon foundries and institutes including HPK [5], FBK [6], CNM [3], BNL [7], NDL [8, 9, 10, 11]. The Institute of High Energy Physics (IHEP) High-Granularity Timing Detector group has recently developed the first version of IHEP-IME LGAD (IHEP-IMEv1) sensors with the Institute of Microelectronics (IME) of the Chinese Academy of Sciences [12], which are aimed to be used as sensors for the HGTD project.

In this paper, the production and pre-irradiation characteristics of these IHEP-IMEv1 sensors are presented. Firstly, the architectures and fabrications of the sensors are detailed with the wafer description of different gain layer implantation parameters. Next, some electrical characteristics such as leakage current behavior (I-V) and capacitance-voltage (C-V) properties are reported. Finally, time resolution and charge collection results obtained from the beta test are shown.

# 2. LGAD sensor structure

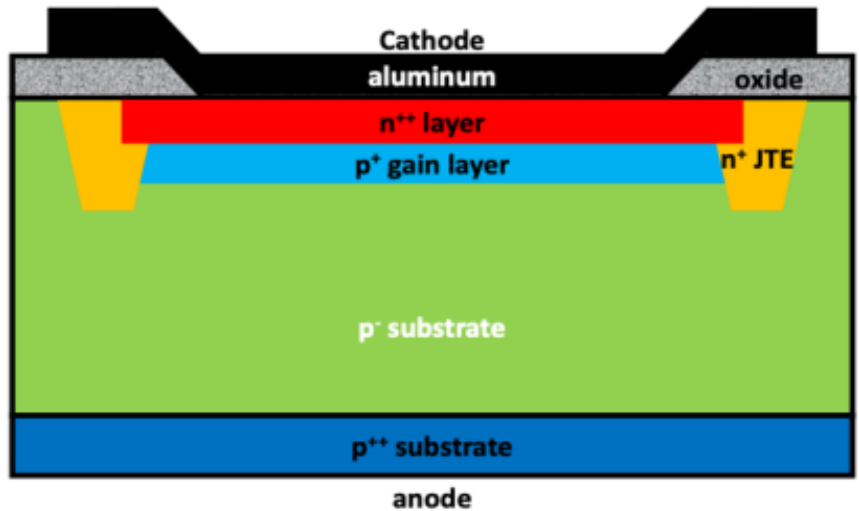

Figure 1: The basic structure of Low Gain Avalanche Detectors (LGAD).[12]

Figure 1 shows the basic structure of Low Gain Avalanche Detectors (LGAD). The most important part of the LGAD sensor is the charge multiplication in the so-called gain layer which is formed by a heavily doped 1–2 μm thick p+ layer sandwiched between the n++ layer and the p− substrate. A further important structure in an LGAD is the JTE (Junction Termination Extension), implanted to avoid early breakdown at the edge. The thickness of the p− substrate is as thin as 50 μm for a superior time resolution of around 35ps per detector layer [16]. The negative anode voltages are applied from the backside electrode to make the sensor work at the voltage before breakdown and with a gain larger than 10 for good time resolution, while the signals of each sensor are collected and analyzed from the top cathode electrode. The measurements in this paper were done with single pad sensors with one guard ring (Figure 2(b)).

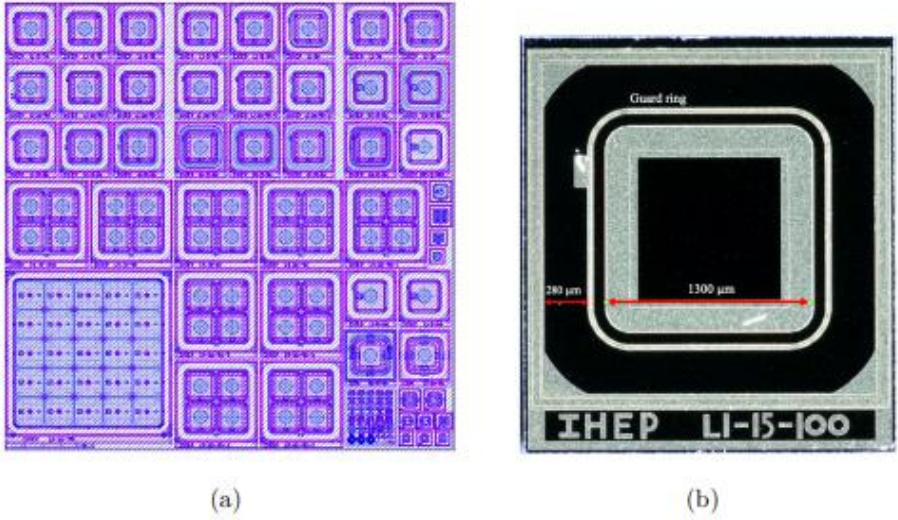

Figure 2: The mask of the IHEP-IMEv1 sensors (a) and the picture of one single pad sensor (b)

## 3. Devices Fabrication

Starting substrates were 8-inch boron-doped (100)-oriented p-type silicon wafers with 50 μm thick EPI (Epitaxial) layer which has a resistivity of 1 kΩ·cm. Three wafers with different process parameters were taped out. Wafer 1 (W1) and Wafer 7 (W7) with 40 keV n++ layer implantation energy, while Wafer 8 (W8) with 50 keV n++ implantation energy. With each wafer having four quadrants, four different doses for p+ layer were implanted into different layer implantation is 400 keV, same for all the quadrants in the three wafers. Compared with W7, W1 has no other different process except carbon implantation to the gain layer. The implantation parameters of the p+ layer and n++ layer for different wafers and quadrants are listed in table 1.

Table 1: Process parameters of p+ layer and n++ layer for different wafers(W1/W7/W8) and quadrants(I/II/III/IV)

| | | W1 | | | | W7 | | | | W8 | | | |
|---|---|---|---|---|---|---|---|---|---|---|---|---|---|
| carbon implantation | | yes | | | | no | | | | no | | | |
| n++ layer | Energy (keV) | 40 | | | | 40 | | | | 50 | | | |
| | Dose ($1\times10^{15}$cm$^{-2}$) | 1 | | | | 1 | | | | 1 | | | |
| p+ layer | Energy (keV) | 400 | | | | 400 | | | | 400 | | | |
| | Quadrant | I | II | III | IV | I | II | III | IV | I | II | III | IV |
| | Dose ($1\times10^{12}$cm$^{-2}$) | 1.5 | 1.9 | 2.2 | 2.5 | 1.5 | 1.9 | 2.2 | 2.5 | 1.5 | 1.9 | 2.2 | 2.5 |

After the dicing of the wafers, leakage current (I-V) and capacitance-voltage (C-V) characteristics of the LGAD sensors were measured using Keithley 2410, Keithley 2400 and Keysight E4980A LCR meter to obtain the electrical properties. For the I-V test, the Keithley 2410 provided high voltage, while the Keithley 2400 measured the leakage current of the pad. For the C-V measurements, the Keithley 2410 provided high voltage, while the LCR meter measured the capacitance. Time resolution and charge collection characteristics of the sensors at room temperature (20°C) and low temperature (-30°C) were measured by bonding the sensors to UCSC testing boards [15] and measuring timing performance with a Beta source ($^{90}$Sr). A NDLv3 sensor[8] with a time resolution 35.0ps at room temperature and 31.2ps at -30°C was used as a reference, the constant fraction discriminator (CFD) method in frontend electronics and data analysis was used to avoid the time walk effect[17].

## 4. Leakage current and capacitance-voltage characteristics

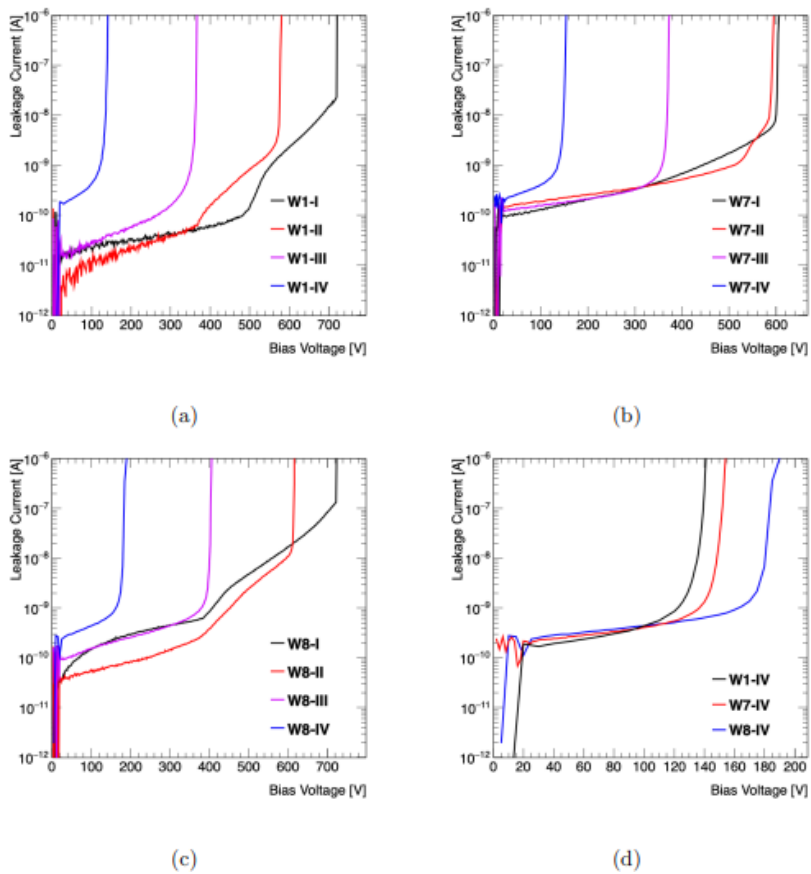

Figure 3: Test results of I-V characteristics for sensors of a) W1, b) W7 and c) W8 with four different p++ layer doses. Slice d) reports a comparison of results for quadrant IV of each wafer.

As we can see from Figure 3, the leakage currents for all sensors with different implantation parameters at room temperature before breakdown are much less than 1nA, which satisfies the requirement of the ATLAS HGTD project (before irradiation). The results in the Figure 3 are the test results from individual sensors. The test results of more than 20 sensors shows that the RMS of the breakdown voltage is about 5 V and the RMS of leakage current is only about 0.1nA. The $V_{BD}$ which is defined as the voltage at which the leakage current reaches 500nA decreases as the dose of the p+ layer goes up. Specifically, the $V_{BD}$ decreases to about 140 V as the dose of p+ layer increases by about $0.2\times10^{12}$cm$^{-2}$(Figure 4). The results fit with TCAD [13] simulation which shows that the electrical field between the p+ layer and n++ layer is higher when increasing the p+ layer dose (Figure 5), so that the $V_{BD}$ decreases. The TCAD simulations were performed both before and after the sensor fabrication. The simulations before sensor fabrication offered a rough range of process parameters. The simulations after fabrication which were calibrated by SIMS result presented precise results compared to sensor test. In the simulation, the "DopingDependence" and "HighFieldSaturation"were used on mobility, "DopingDependence" and "TempDependence" were used on Shockley–Read–Hall Recombination (SRH) and the van Overstraeten de Man model were used for the avalanche mode[18]. The active area which is the region between the p+ and the n++ implantation electrical field of sensors from the first and second quadrant which have lower gain layer dose is not very high, so the edge structures should have a larger effect on the breakdown than the p+ dose, which could explain why the first two quadrants' breakdown voltages of W7 are relatively close.

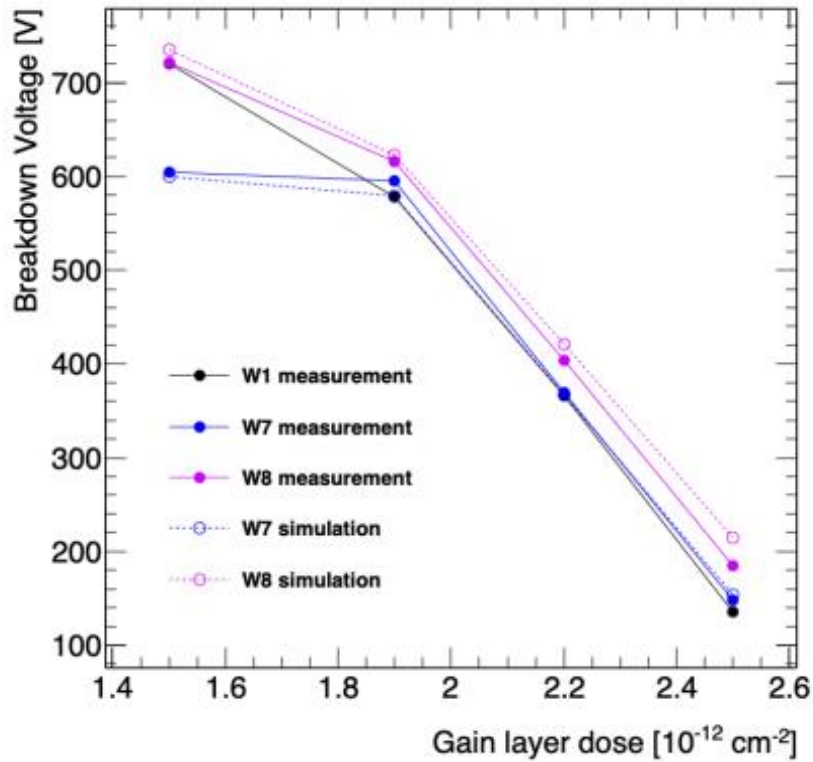

Figure 4: The breakdownn voltage change with the gain layer dose

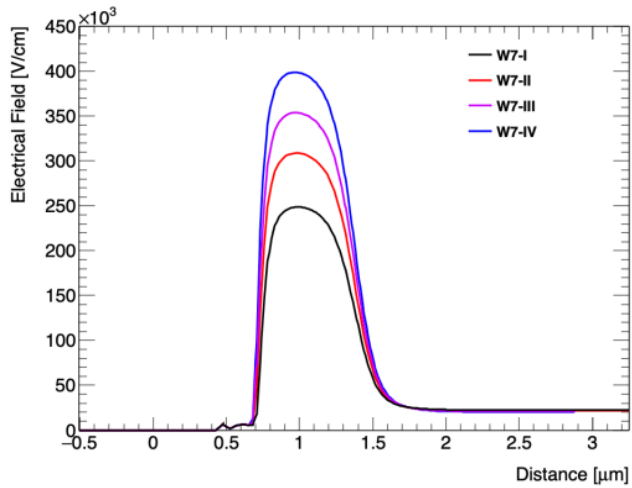

Figure 5: Simulation results of electrical field characteristics for sensors of W7 with four different p++ layer doses (backside voltage -100 V)

Compared with the sensors of the fourth quadrant of wafer 7, sensors from the same quadrant of wafer 8 with higher n++ implantation energy have higher breakdown voltages, about 50 V higher. The doping profiles from SIMS test results shown on Figure 6 revealed that the n++ layer of wafer 8 which uses higher implantation energy goes deeper into the substrate from the surface and so that the distance between the n++ layer and peak of p+ layer becomes narrower, while the two have the same conditions of p+ layer implantation and n++ layer dose. Then we can get a conclusion that even if the doses of the p+ layer and n++ layer are the same, the relative position of the p+ layer and n++ layer will also affect the $V_{BD}$. Wafer 1 with carbon implantation shows a little lower breakdown voltage and a similar leakage current compared with wafer 7 with the same gain layer implantation. The results in the Figure 7 are the test results from individual sensors. The test results of more than 20 sensors shows that the RMS of the $V_{GL}$ is about 0.1V in Wafer 7 and 0.2 V in Wafer 1.

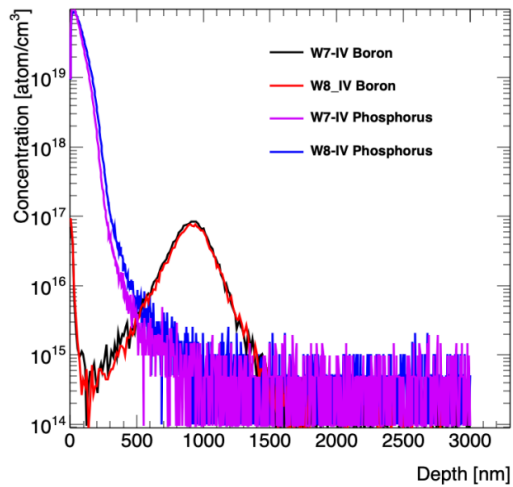

Figure 6: SIMS results of doping profiles for p+ layer(boron) and n++ layer (Phosphorus) of wafer 7-IV and wafer 8-IV

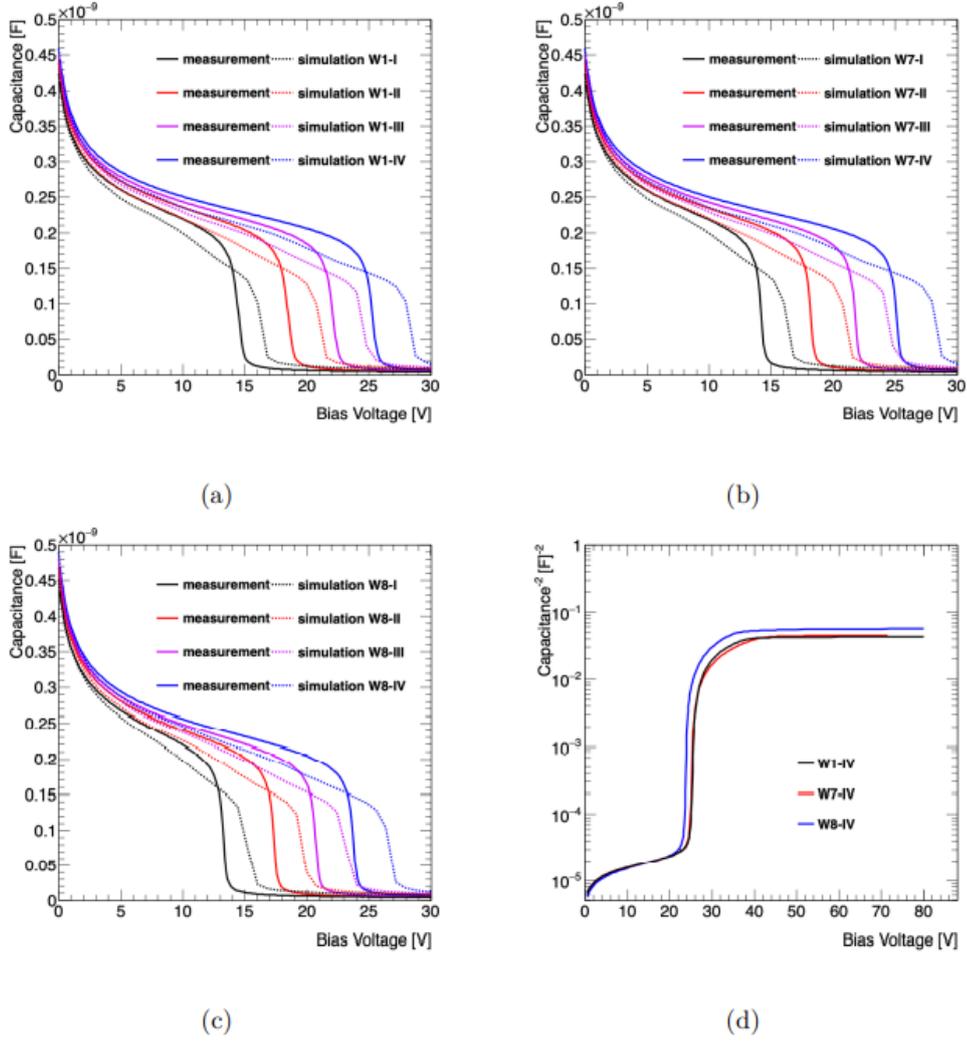

Figure 7: Simulation results and testing results about C-V characteristics for sensors in a)W1, b) W7, and c) W8 with four different p++ layer doses. Slice d) reports a comparison of $1/C^2$-V characteristics for quadrant IV of each wafer.

Gain layer depletion voltage ($V_{GL}$) increases with the p+ layer dose increasing. The trends of the capacitance properties also fit with TCAD simulation results, while $V_{GL}$ from testing results are about 3-5 V higher than simulation results.

Wafer 1 with carbon implantation shows similar capacitance properties compared with wafer 7 with the same gain layer implantation. And for wafer 1 and wafer 7, the capacitance at depletion voltage is about 4.8 pF and the $V_{GL}$ is 24.3 V. But for wafer 8, the capacitance at depletion voltage is 4.25 pF and the $V_{GL}$ is 23.1 V, which means that sensors from wafer 8 with higher n++ implantation energy show lower capacitance at depletion voltage and lower $V_{GL}$. Considering the doping profiles, we can get a conclusion that even if the doses of the p+ layer and n++ layer are the same, the relative position of the p+ layer and n++ layer will also affect the capacitance at depletion voltage and the $V_{GL}$. Keeping the same p+ layer implantation parameters 10 and n++ layer dose, by using higher energy for n++ layer implantation, we can increase the $V_{BD}$ for LGAD sensors but decrease the $V_{GL}$.

## 5. Time resolution and charge collection

The sensors from the fourth quadrants of the three wafers with the highest p+ layer dose shows the lowest breakdown voltage and highest gain layer voltage, which means that these sensors can work at the lowest voltage and can have good properties after irradiation, so they were chosen for testing of time resolution and charge collection properties before irradiation. The W1-IV, W7-IV, W8-IV sensors are wire bonded to the testing board designed by UCSC [14]. Then the beta source ($^{90}$Sr) was used as the excitation source of the sensors.

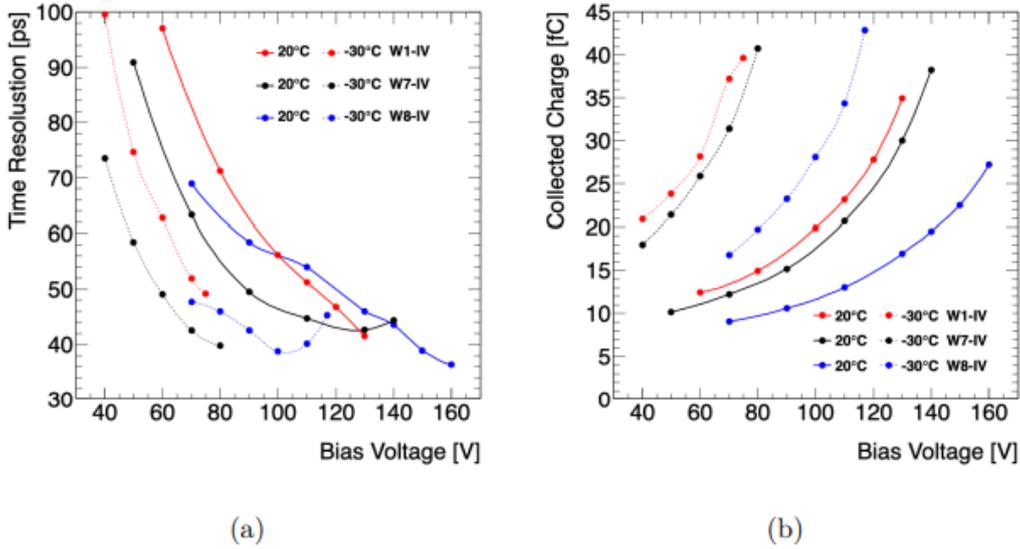

Figure 8: Time resolution (a) and collected charge (b) as a function of bias voltage for W1-IV, W7-IV and W8-IV before irradiation at 20 °C and -30 °C

Figure 8 (a) shows the time resolution as a function of bias voltage for IHEP-IMEv1 LGAD sensors measured at 20°C and -30°C before irradiation. The measurement shows that the time resolution of the sensors of W1-IV, W7-IV, and W8-IV, which have p+ layer dose up to $2.5\times10^{12}$ cm$^{-2}$ can be lower than 40ps at 20°C and -30°C. These three sensors have the potential to satisfy the ATLAS physics requirement in the HGTD project (< 35ps before irradiation). The voltages for time resolution better than 40ps of the three sensors at -30 °C are all lower than that at 20°C, which means that the three LGAD sensors can work at a lower voltage for working temperature as -30°C than 20°C. While W8 with higher n++ implantation energy and higher breakdown voltage shows higher operation voltage for same time resolution. Figure 8 (b) shows the collected charges as a function of the bias voltage for IHEP-IMEv1 LGAD sensors before irradiation at 20°C and -30°C. When high voltage is applied, IHEP-IMEv1 LGAD sensors have a collected charge larger than 15fC before irradiation, some of them can reach 30fC, which satisfies the requirement of the ATLAS HGTD project (>15fC before irradiation). The voltages for the collected charge larger than 15fC of the three sensors at -30°C are all lower than the results at 20°C, which also means that the three LGAD sensors can work at a lower voltage at low temperature. While W8 with higher n++ implantation energy and higher breakdown voltage also shows higher operation voltage for the same charge collection. W1 with carbon implantation shows a slightly operation voltage for same charge collection at -30℃.

From all the results above, the IHEP-IMEv1 satisfies the requirement of the ATLAS HGTD project before irradiation in terms of breakdown voltage, gain layer voltage, time resolution, and charge collection. W1-IV, W7-IV and W8-IV were sent to be irradiated and testing results for sensors post irradiation will be analyzed next. Furthermore, in the IHEP-IMEv2 run, IHEP and IME will add large array sensors (15×15) and optimize the condition of carbon implantation to improve the performance of charge collection after irradiation.

## 6. Summary

In summary, the LGAD devices developed by IHEP and fabricated by IME have a time resolution better than 35 ps while the charge collection is better than 15 fC at 20℃ and -30℃. By comparing sensors with different implantation conditions, it can be found that increasing p+ layer dose can lead to the VBD decreasing and the VGL increasing. While deeper n++ layer can lead to the VBD increasing but the VGL decreasing. With or without carbon implantation, sensors show similar properties before irradiation. All the test results show that the IHEP-IMEv1 sensors fulfill the specification of ATLAS HGTD sensors' requirements before irradiation. IHEP-IMEv1 sensors were sent to be irradiated and test results about sensors post-irradiation performance will be subject to a later paper.

## Acknowledgment

This work was supported by the National Natural Science Foundation of China(No.11961141014), the State Key Laboratory of Particle Detection and Electronics, China, SKLPDEZZ-201911 project and SKLPDE-ZZ202001 project, and the Scientific Instrument Developing Project of the Chinese Academy of Sciences, Grant No. ZDKYYQ20200007.